\documentclass[aps,preprint,superscriptaddress]{revtex4}
\usepackage{graphicx}
\usepackage{subfigure}
\usepackage{float}
\usepackage{amsmath}
\usepackage{amsfonts}
\usepackage{color}
\usepackage{bm}
\usepackage{bbm}
\usepackage{txfonts}
\usepackage{array}
\usepackage{amssymb}
\usepackage{hyperref}
\begin{document}
\title{Effect of super-Gaussian pulse shape on pair production in chirped electric field with spatial inhomogeneity}
\author{Xiao-Ting Xu}
\affiliation{College of Physics and Electronic Engineering, Northwest Normal University, Lanzhou 730070, China}
\affiliation{School of Mathematics and Physics, Lanzhou Jiaotong University, Lanzhou 730070, China}
\author{Le-Le Chen}
\affiliation{College of Physics and Electronic Engineering, Northwest Normal University, Lanzhou 730070, China}
\affiliation{School of Mathematics and Physics, Lanzhou Jiaotong University, Lanzhou 730070, China}
\author{Rong-an Tang}
\affiliation{College of Physics and Electronic Engineering, Northwest Normal University, Lanzhou 730070, China}
\author{Xue-Ren Hong \footnote{hongxr@nwnu.edu.cn.}}
\affiliation{College of Physics and Electronic Engineering, Northwest Normal University, Lanzhou 730070, China}
\author{Lie-Juan Li \footnote{ljli@lzjtu.edu.cn.}}
\affiliation{School of Mathematics and Physics, Lanzhou Jiaotong University, Lanzhou 730070, China}
\author{Bai-Song Xie \footnote{bsxie@bnu.edu.cn}}
\affiliation{Key Laboratory of Beam Technology of the Ministry of Education, and School of Physics and Astronomy, Beijing Normal University, Beijing 100875, China}
\affiliation{Institute of Radiation Technology, Beijing Academy of Science and Technology, Beijing 100875, China}
\date{\today}

\begin{abstract}
Pair production in spatially inhomogeneous chirped electric fields with super-Gaussian pulse shape is investigated using the Dirac-Heisenberg-Wigner formalism, and the effect of super-Gaussian pulse shapes on the reduced momentum spectrum and the reduced total yield of created particles is mainly concerned.
It is found that with the variation of the super-Gaussian envelope exponent, the momentum spectrum exhibits the more pronounced oscillations, shifting and broadening.
The total yield of created particles increases monotonically with the increase of the super-Gaussian envelope exponent in the high-frequency fields with small chirp and low-frequency fields with any chirp.
Meanwhile, the total yield of created particles under the super-Gaussian pulse electric fields is approximately twice that produced with the usual Gaussian pulse envelope. These results can provide theoretical guidance for optimizing the form of external field to enhance the vacuum pair production rate.

\end{abstract}
\maketitle

\section{Introduction}

Electron-posirton $({e^{+}e^{-}})$ pair production in vacuum under intense electromagnetic fields represents one of the most captivating phenomena in relativistic quantum physics \cite{dirac1928quantum,sauter1931verhalten,schwinger1951gauge,Gelis:2015kya,DiPiazza:2011tq,xie2017electron}.
Dirac was the first to formulate the relativistic wave equation for electrons and theoretically predict the existence of the positron \cite{dirac1928quantum}.
Sauter discovered that ${e^{+}e^{-}}$ pairs can be created from vacuum by the tunneling mechanism when studying the Dirac equation in a constant electric field \cite{sauter1931verhalten}.
Subsequently, Schwinger studied pair production in a strong static electric field using the proper-time method and derived the critical electric field strength $E_\text{cr}=m^{2} c^{3} / e \hbar \approx 1.3 \times 10 ^{16} \rm {V/cm}$, corresponding to a laser intensity of $ I_\text{cr} \approx 4.3 \times 10^{29} \rm W / \rm cm^{2}$ (where $m$ and $-e$ are the electron mass and charge, respectively) \cite{schwinger1951gauge}.
Since then, the pair production from the vacuum in intense external fields is also known as the Sauter-Schwinger effect \cite{Gelis:2015kya,DiPiazza:2011tq,xie2017electron}.
However, the current laser intensity reaches only approximately $10 ^{23} \rm W /\rm cm^{2}$, which is much lower than the critical electric field strength \cite{2021Ultra}. So it is not yet possible to create observable ${e^{+}e^{-}}$ pair in experiments.
Fortunately, the development of ultraintense laser technology might make the experimental observation of pair production possible in the near future \cite{Heinzl:2008an, Marklund:2008gj, Pike:2014wha}.
Especially the application of chirped pulse amplification (CPA) technique has greatly improved the laser intensity, and the expected intensity to be $10 ^{25} \sim 10 ^ {26}\rm W /\rm cm^{2}$ for many planned facilities \cite{Strickland:1985gxr, Dunne:2008kc}.
Moreover, the already operating X-ray free electron laser (XFEL) is expected to achieve subcritical field strengths of ${E/{E}_{\mathrm{cr}}\approx 0.01 \sim 0.1}$ \cite{Ringwald:2001ib}.

It is well known that the pair production from the vacuum in strong external fields has been investigated using a variety of theoretical approaches \cite{WKB:2010Dumlu,WKB:2022CK,WKB:2021HT,Dumlu:2011rr,Gies:2005bz,Schneider:2014mla,Kluger:1991ib, Alkofer:2001ik, Abdukerim:2017hkh,liliejuan:2025qve,Kohlfurst:2015zxi, Xie:re2017,cnzPRD109,Kohlf2020Effect}, and the commonly used methods can be roughly divided into two categories. One is semiclassical approximation methods, such as the Wentzel-Kramers-Brillouin (WKB) approximation \cite{WKB:2010Dumlu,WKB:2022CK,WKB:2021HT,Dumlu:2011rr} and the worldline instanton technique \cite{Gies:2005bz,Schneider:2014mla}. The other is quantum kinetic theory, including the quantum Vlasov equation (QVE) \cite{Kluger:1991ib, Alkofer:2001ik, Abdukerim:2017hkh,liliejuan:2025qve} and Dirac-Heisenberg-Wigner (DHW) formalism \cite{ Kohlfurst:2015zxi, Xie:re2017,cnzPRD109,Kohlf2020Effect}.
Currently, the quantum kinetic theory is widely employed, as it can obtain important phase-space information of the created particles such as momentum spectrum, which is more conducive to understanding the pair production process.
For example,
Hebenstreit \textsl{et al}. have investigated Schwinger pair production in short laser pulses with a subcycle structure using the QVE solution method and demonstrated that the momentum spectrum of the generated particles is highly sensitive to the external field parameters \cite{Hebenstreit:2009km}. Subsequently, they have studied the pair production in a simple space- and time-dependent electric fields pulse employing the DHW formalism and predicted a self-bunching effect of particles in phase space \cite{Hebenstreit:2011wk}.
Kohlf\"urst \textsl{et al}. have explored the pair production process in the time-oscillating and spatially inhomogeneous electric field employing the DHW formalism and revealed that the ponderomotive effect is significant in multiphoton pair production \cite {Kohlfurst:2017hbd}.
Ababekri \textsl{et al}. have studied the pair production from vacuum in the space- and time-dependent external electric fields using DHW formalism and found that the finite spatial scale has nontrivial effects on Schwinger pair production \cite{Ababekri:2019dkl}.
Sch\"{u}tzhold \textsl{et al}. have investigated the pair production in a low-frequency strong field superimposed with a high-frequency weak field and found that the dynamically assisted Schwinger mechanism can significantly enhance the particle production rate \cite{LiLieJuan2021,Schutzhold:2008pz,HTaya:2020PRR,CK:2022PRR}.
Moreover, numerous researches on pair production in time-dependent electric fields and spatially inhomogeneous oscillating electric fields have found that the frequency chirp affects the momentum spectrum and total yield of the created particles \cite{Dumlu:2010vv,Olugh:2018seh,bai2022enhancement}.
Also, Nuriman \textit{et al.} have investigated the effect of super-Gaussian pulse shape on pair production in a spatially homogeneous external field using the QVE solution method and found that the flat-top super-Gaussian pulse shape has an advantage in pair production \cite{Nuriman:2012super}.

At present, the influence of super-Gaussian pulse shapes on vacuum pair production has been limited to the spatially homogeneous electric fields, but the spatial inhomogeneity of the external field has not yet been considered.
In this paper, the effect of super-Gaussian pulse shape on pair production is investigated in chirped electric fields with spatial inhomogeneity by using the DHW formalism.
The momentum spectrum and the total yield of the created particles are studied in high- and low- frequency fields with different super-Gaussian envelope exponents, and some optimized parameters for pair production are obtained.
Throughout this paper, the natural units $\hbar=c=1$ are used and all quantities are expressed in terms of the electron mass $\mathrm{m}$.

The paper is structured as follows. A brief overview of the DHW formalism applied in this study is provided in Sec. \ref{DHWformalism}. The model of background fields is introduced in Sec. \ref{field}. The numerical results for the momentum spectrum and the total yield of the created particles are presented in Sec. \ref{momentum}. The semiclassical analysis and discussion about the numerical results is shown in Sec. \ref{turningpoint}. Finally, a conclusion and outlook is presented in Sec. \ref{conclusion}.

\section{Theoretical formalism: DHW formalism}\label{DHWformalism}

The DHW formalism is a quantum kinetic theory whose central idea is to employ the Wigner function to describe the relativistic phase-space distribution \cite{Vasak:1987um,Phase-space str}. This formalism has been widely used to investigate vacuum pair production in arbitrary electromagnetic fields \cite{Hebenstreit:2011wk,Kohlfurst:2017hbd,Ababekri:2019dkl,LiLieJuan2021,Ali2025}.
Since the detailed derivations of the DHW formalism have been presented in previous works \cite{Phase-space str,Kohlfurst:2015zxi},
only the basic ideas and essential points of this method are briefly outlined here.

We begin with the gauge-covariant density operator $\hat {\mathcal C}_{\alpha \beta} \left(r , s \right)$ of the system based on
\begin{equation}\label{DensityOperator}
 \hat {\mathcal C}_{\alpha \beta} \left(r , s \right) = \mathcal U \left(A,r,s
\right) \ \left[ \bar \Psi_\beta \left( r - s/2 \right), \Psi_\alpha \left( r +
s/2 \right) \right],
\end{equation}
where $\Psi$ represents Dirac field operator, $r$ represents the center-of-mass coordinate and $s$ is the relative coordinate \cite{Phase-space str}.
The Wilson line factor $\mathcal U \left(A,r,s \right)$ before the commutator is a factor that ensures the gauge invariance of the density operator as
\begin{equation}\label{Wilson line factor}
 \mathcal U \left(A,r,s \right) = \exp \left[ \mathrm{i} e s \int_{-1/2}^{1/2} \mathrm{d} \xi \ A \left(r+ \xi s \right)  \right].
\end{equation}
Obviously, it is related to the elementary charge $e$ and the background gauge field $A$.

The covariant Wigner operator $\hat{\mathcal W}_{\alpha \beta} \left( r , p \right)$ is obtained via the Fourier transform of the covariant density operator $\hat {\mathcal C}_{\alpha \beta} \left(r , s \right)$ from the $s$-space to the $p$-space as
\begin{equation}\label{WignerOperator}
 \hat{\mathcal W}_{\alpha \beta} \left( r , p \right) = \frac{1}{2} \int \mathrm{d}^4 s \
\mathrm{e}^{\mathrm{i} ps} \  \hat{\mathcal C}_{\alpha \beta} \left( r , s \right).
\end{equation}
Considering the vacuum expectation value of the covariant Wigner operator $\hat{\mathcal W}_{\alpha \beta} \left( r , p \right)$, the Wigner function $\mathbbm{W} \left( r,p \right)$ can be defined as
\begin{equation}\label{Wigner function}
 \mathbbm{W} \left( r,p \right) = \langle \Phi \vert \hat{\mathcal W}_{\alpha \beta} \left( r,p
\right) \vert \Phi \rangle.
\end{equation}
For the convenience of numerical calculations, the Wigner function $\mathbbm{W} \left( r,p \right)$ is decomposed into 16 covariant Wigner components using a complete basis set $\{ \mathbbm{1}, \gamma_5, \gamma^\mu, \gamma^\mu \gamma_5, \sigma^{\mu\nu}=:\frac{i}{2} [\gamma^\mu, \gamma^\nu]\}$ as
\begin{equation}\label{decomposed}
\mathbbm{W} \left( r,p \right) = \frac{1}{4} \left( \mathbbm{1} \mathbbm{S} + \textrm{i} \gamma_5
\mathbbm{P} + \gamma^{\mu} \mathbbm{V}_{\mu} + \gamma^{\mu} \gamma_5
\mathbbm{A}_{\mu} + \sigma^{\mu \nu} \mathbbm{T}_{\mu \nu} \right) ,
\end{equation}
where $\mathbbm{S}$, $\mathbbm{P}$, $\mathbbm{V}_{\mu}$, $\mathbbm{A}_{\mu}$ and $\mathbbm{T}_{\mu \nu}$ denote scalar, pseudoscalar, vector, axial vector and tension, respectively.

However, it is found that to numerically solve the equation of motion for the covariant Wigner function at all spacetime points is nearly an impossible task. Thus, a convenient and effective approach is to obtain the equal-time Wigner function $\mathbbm{w} \left( \mathbf{x}, \mathbf{p}, t \right)$ by taking the energy average of the covariant Wigner function $\mathbbm{W} \left( r,p \right)$ as
\begin{align}
 \mathbbm{w} \left( \mathbf{x}, \mathbf{p}, t \right) = \int \frac{\mathrm{d} p_0}{2 \pi}
\ \mathbbm{W} \left( r,p \right).
\end{align}

In our investigation of $1+1$ dimensional spatially inhomogeneous electric field, one finds that only
$\mathbbm{S}$, $\mathbbm{P}$, and $\mathbbm{V}_{\mu} (\mu=0,1)$ are nonzero, and the corresponding physical quantities are denoted by lowercase in the equal-time case. Hence, the equations of motion can be simplified to the following $4$ equations as \cite{Hebenstreit:2011wk,Kohlfurst:2015zxi}
\begin{align}
 &D_t \mathbbm{s} - 2 p_x \mathbbm{p} = 0 , \label{pde:1}\\
 &D_t \mathbbm{v}_{0} + \partial _{x} \mathbbm{v}_{x} = 0 , \label{pde:2}\\
 &D_t \mathbbm{v}_{x} + \partial _{x} \mathbbm{v}_{0} = -2 m \mathbbm{p} , \label{pde:3}\\
 &D_t \mathbbm{p} + 2 p_x \mathbbm{s} = 2 m \mathbbm{v}_{x} , \label{pde:4}
\end{align}
where $D_t$ is the differential operator as
\begin{equation}\label{pseudoDiff}
 D_t = \partial_{t} + e \int_{-1/2}^{1/2} \mathrm{d} \xi \,\,\, E_{x} \left( x + i \xi \partial_{p_{x}} \, , t \right) \partial_{p_{x}} .
\end{equation}
Meanwhile, the corresponding vacuum initial conditions of the Eqs.(\ref{pde:1})-(\ref{pde:4}) can be given by
\begin{equation}\label{vacuum-initial}
{\mathbbm s}_{vac} = - \frac{2m}{\Omega} \, ,
\quad {\mathbbm v}_{vac} = - \frac{2{ p_x} }{\Omega} \,  ,
\end{equation}
where $\Omega=\sqrt{p_{x}^{2}+m^2}$ is the energy of a particle \cite{Kohlfurst:2015zxi}.
By subtracting these vacuum terms, the modified Wigner component $\mathbbm{w}_{k}^{v}\left( x , p_{x} , t \right)$ can be written as
\begin{equation}\label{NS}
\mathbbm{w}_{k}^{v}\left( x , p_{x} , t \right)=\mathbbm {w}_{k}\left( x , p_{x} , t \right) - {\mathbbm w}_{vac}\left( p_{x} \right),
\end{equation}
where $\mathbbm{w}_{k}$ is the Wigner component in Eqs.(\ref{pde:1})-(\ref{pde:4}) with the correspondence: $\mathbbm{w}_{0}={\mathbbm s}$, $\mathbbm{w}_{1} = \mathbbm {v}_{0}$, $\mathbbm{w}_{2} = \mathbbm {v}_{x}$ and $\mathbbm{w}_{3} = \mathbbm {p}$, and $\mathbbm {w}_{vac}$ denotes the corresponding vacuum initial condition given in Eq.(\ref{vacuum-initial}).

Then, the particle number density $n \left( x , p_{x} , t \right)$ is given by \cite{Hebenstreit:2011wk,Kohlfurst:2017hbd}
\begin{equation}\label{PS}
n \left( x , p_{x} , t \right) = \frac{m \mathbbm{s}^{v} \left( x , p_{x} , t \right) + p_{x}  \mathbbm{v}_{x}^{v} \left( x , p_{x} , t \right)}{\Omega \left( p_{x} \right)},
\end{equation}
and particles number density in the momentum space could be obtained  from $n \left( x , p_{x} , t \right)$  by integrating out $x$ as
\begin{equation}\label{MS}
n \left( p_{x} , t \right) = \int \mathrm{d} x \, n \left( x , p_{x} , t \right).
\end{equation}
Accordingly, the total yield of the created particles $N\left(t \right)$ can be calculated by
\begin{equation}\label{Num}
N\left(t \right) = \int \mathrm{d}p_x\, n \left( p_{x} , t \right).
\end{equation}
To obtain the nontrivial dependence on the spatial scale $\lambda$, we calculate the reduced quantities $\bar n \left( p_{x} , t \right) \equiv n \left( p_{x} , t \right) / \lambda$ and $\bar N\left(t\rightarrow \infty \right) \equiv N \left(t \rightarrow \infty \right) /\lambda$.

\section{Model of the background electric field}\label{field}

We investigate $e^{+}e^{-}$ pair production in 1+1 dimensions for super-Gaussian pulse shape effects in chirped electric field with spatial inhomogeneity.
The external field model can be regarded as an ideal standing wave model, consisting of a spatially dependent electric field constituted by two counter-propagating coherent laser fields, where the magnetic field components cancel each other out \cite{Ababekri:2019qiw}. The field direction is along the $x$-axis, and its strength varies with $x$ and $t$. Then, the external field model expression is given by
\begin{equation}\label{FieldMode}
\begin{aligned}
E\left(x,t\right)=E_{0}\exp \left(-\frac{x^{2}}{2 \lambda^{2}} \right )\exp \left[-\frac{1}{2}\left(\frac{t}{\tau}\right)^{2l} \right ]\cos(\omega t+b t^{2}),
\end{aligned}
\end{equation}
where $E_{0}$ denotes the strength of the electric field, the integer $l$ is super-Gaussian envelope exponents, $\omega$ is the oscillating frequency of the fields, $\lambda$ and $\tau$ are the characteristic time and spatial scale, $b$ represents the corresponding frequency chirp, and the effective time-dependent frequency can be introduced by $\omega_{\mathrm{eff}} = \omega+ 2bt$. Under this external field, it is assumed that the particles are predominantly created in the field direction, and set $p_{\perp} = 0$ \cite{Hebenstreit:2011wk}.

In our study of ${e^{+}e^{-}}$ pair production, the physical parameters selected are as follows. $E_{0}=0.5E_{\mathrm{cr}}$ is chosen and $l=1,2,5$ is considered.
For the high-frequency electric fields, $\omega= 0.7~\mathrm{m}$ and $\tau = 45~\mathrm{m}^{-1}$ are taken, and for the low-frequency electric fields, $\omega= 0.1~ \mathrm{m}$ and $\tau =25~\mathrm{m}^{-1}$ are selected.
The chirp parameter $b$ is expressed as $b= \alpha \omega/2\tau$ $(\alpha \geq 0)$ to characterize the variation of the effective frequency induced by the chirp.
For the high-frequency fields, the upper limit of the chirp value $b= 0.5\omega/\tau \approx 0.0078~\mathrm{m}^{2}$ is set by keeping the maximum effective frequency to be around the threshold frequency $\omega_{\mathrm{eff}}(\tau) \sim 1.0~\mathrm{m}$.
For the low-frequency fields, the maximum chirp value $b= 1.5\omega/\tau = 0.006~\mathrm{m}^{2}$ is taken as $\omega_{\mathrm{eff}}(\tau)\tau \sim \mathcal{O}(1)$ to maintain the few-cycle pulse shape of the external field \cite{Ababekri:2019qiw,mohamedsedik2021schwinger}.

\section{Numerical results and discussion }\label{momentum}

\subsection{High Frequency Fields: $\omega=0.7~\mathrm{m}$}

In this subsection, it considers the influence of the super-Gaussian envelope exponent $l$ on pair production in the high-frequency electric fields, and discusses the results for three spatial scales: $\lambda= 1000~\mathrm{m}^{-1}$, $\lambda= 10~\mathrm{m}^{-1}$, and $\lambda= 2.5~\mathrm{m}^{-1}$.
\begin{figure}[t]
\begin{center}
\includegraphics[width=\textwidth]{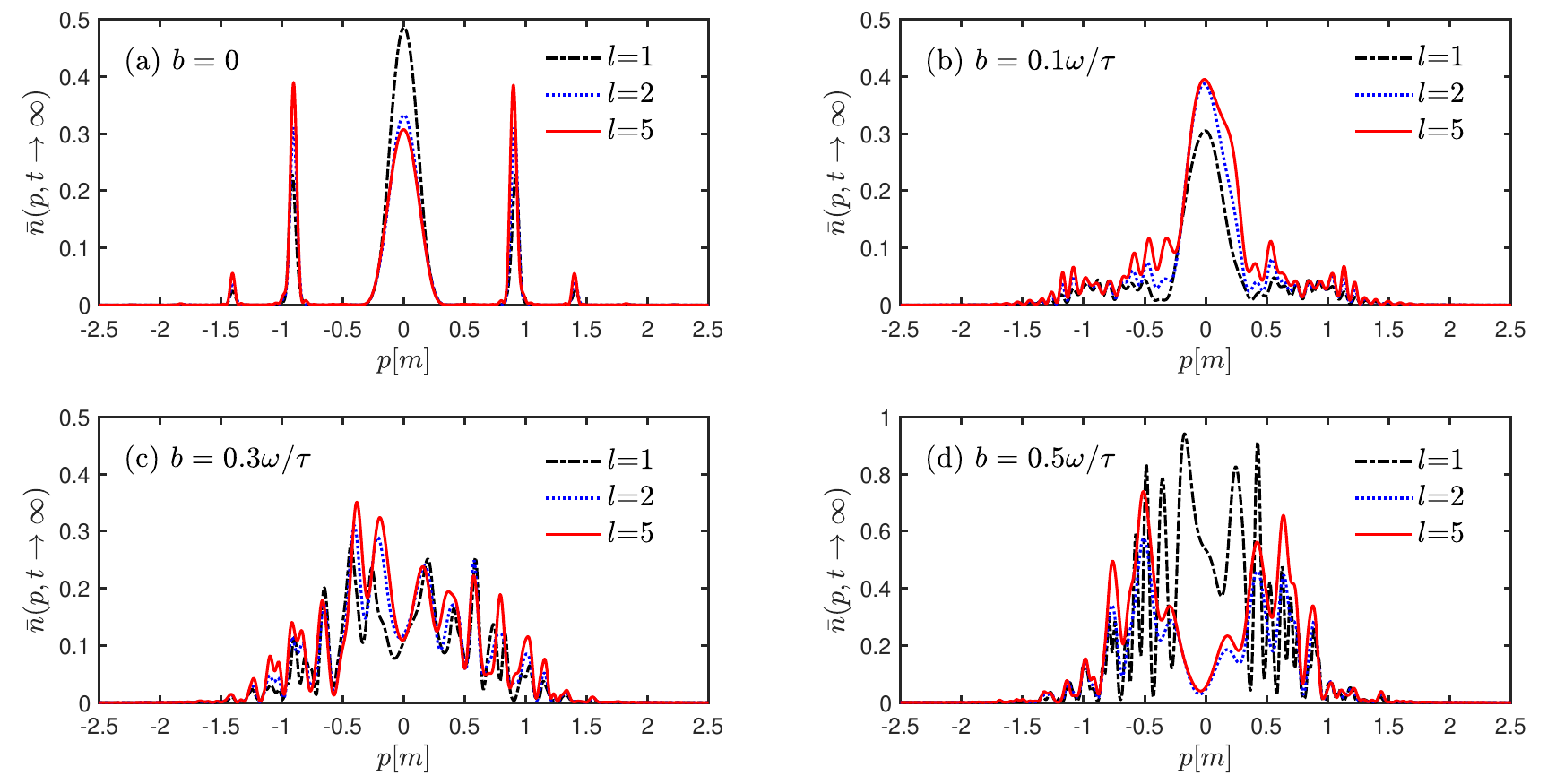}
\end{center}
\setlength{\abovecaptionskip}{-0.5cm}
\caption{Reduced momentum spectrum for different super-Gaussian envelope exponents in the high-frequency electric fields with $\lambda= 1000~\mathrm{m}^{-1}$. The other field parameters are $E_{0}=0.5E_\mathrm{cr}$, $\omega=0.7~\mathrm{m}$ and $\tau=45~\mathrm{m}^{-1}$.}
\label{fig:1}
\end{figure}

Firstly, the momentum spectrum for the quasi-homogeneous limit $\lambda =1000 ~\mathrm{m}^{-1}$ is presented in Fig. \ref{fig:1}.
It is found that for vanishing chirp $b=0$, the momentum spectrum displays typical multiphoton-dominated particle creation, meaning that the characteristic multiphoton peaks are easily distinguishable, see Fig. \ref{fig:1}(a).
This phenomenon can be explained by considering the vector potential being treated as quasi-homogeneous when the external field is on a large spatial scale, allowing the spatial dependence in Eq.(\ref{FieldMode}) to be neglected, and the particle momenta can be estimated as \cite{Kohlfurst:2017hbd}
\begin{equation}\label{particlemomenta}
\begin{aligned}
p_{x,n} \approx \sqrt{\left( \frac{n\omega}{2} \right)^2 - m^2 \left( 1 + \frac{\varepsilon^2 m^2}{2\omega^2} \right)}.
\end{aligned}
\end{equation}
From Eq.(\ref{particlemomenta}), it can be easily inferred that the peak in the middle at $p_x =0$ is associated with the $3$-photon process, while the side maxima at $p_x =0.9~\mathrm{m}$ and $p_x =1.4~\mathrm{m}$ are related to $4$- and $5$-photon pair production, respectively. This result is identical to the results of Fig. 1 in the Ref. \cite{Ababekri:2019qiw}.
Then, when $l=2$ compared with the results for $l=1$ the peak of momentum spectrum in the middle decreases while that of the side peak increases. This phenomenon becomes more pronounced at $l=5$.
This is because the temporal envelope of the external field becomes flatter at the top and steeper at the edges as $l$ increases, which extends the duration of interaction at higher external field intensity and shortens the effective interaction time of the entire pair production process.
It is well known that the energy accumulation condition for multiphoton processes is sensitive to both electric field intensity and time.
Specifically, for $l=5$, the stronger electric field leads to higher particle velocities and thus larger momenta, so the particles are pushed apart more rapidly. As a result, the momentum spectrum exhibits a higher peak in the large-momentum region, indicating a greater accumulation of particles there. In contrast, the external field acts for a longer duration and the particle velocity decreases gradually for $l=1$, $\textit{i.e.}$, resulting in smaller momenta. Consequently, more particles will focus around $p_x \approx 0$, which manifests as a higher peak at $p_x \approx 0$ compared with the cases of $l=2$ and $l=5$.
It is found that the multiphoton peaks shown in Fig. \ref{fig:1}(a) are no longer straightforward to identify and give rise to oscillations, due to the introduction of the chirp, which allows more frequency modes $\omega_{\mathrm{eff}}$ to enter the pair production process, as shown in Figs. \ref{fig:1}(b)-\ref{fig:1}(d).
For small chirp $b=0.1\omega/\tau$, only the peak at $p_x =0$ remains, while other multiphoton peaks are replaced by oscillations, see Fig. \ref{fig:1}(b).
The oscillations can be understood as an interference effect between temporally separated reflected waves due to multiple bump structure of the scattering potential \cite{Dumlu:2010vv}.
When $l=2$, compared with the results of $l=1$, the peak of the momentum spectrum at $p_x =0$ increases and the oscillation becomes more evident.
This can be explained that the small chirp $b=0.1\omega/\tau$ causes the instantaneous frequency to vary with time, increasing the phase accumulation difference for particles generated at different moments. This enhances the interference effect, leading to more pronounced oscillations.
This trend is further pronounced when $l=5$.
As the chirp increases further, strong oscillations emerge in the momentum spectrum, and the modulation effect for momentum spectra by the super-Gaussian envelope exponent $l$ becomes weak, see Figs. \ref{fig:1}(c) and \ref{fig:1}(d).
This is because when the chirp is sufficiently large, the frequency variation dominates the evolution of the momentum spectrum of created particles, while the modulation effect of the temporal envelope is suppressed.
Moreover, the particle number density at $p_x =0$ is also observed to be larger for $l=1$ than for $l=2$ and $l=5$ in Fig. \ref{fig:1}(d).
A detailed explanation of this behavior will be presented in Sec. \ref{turningpoint} using the semiclassical analysis.
\begin{figure}[t]
\begin{center}
\includegraphics[width=\textwidth]{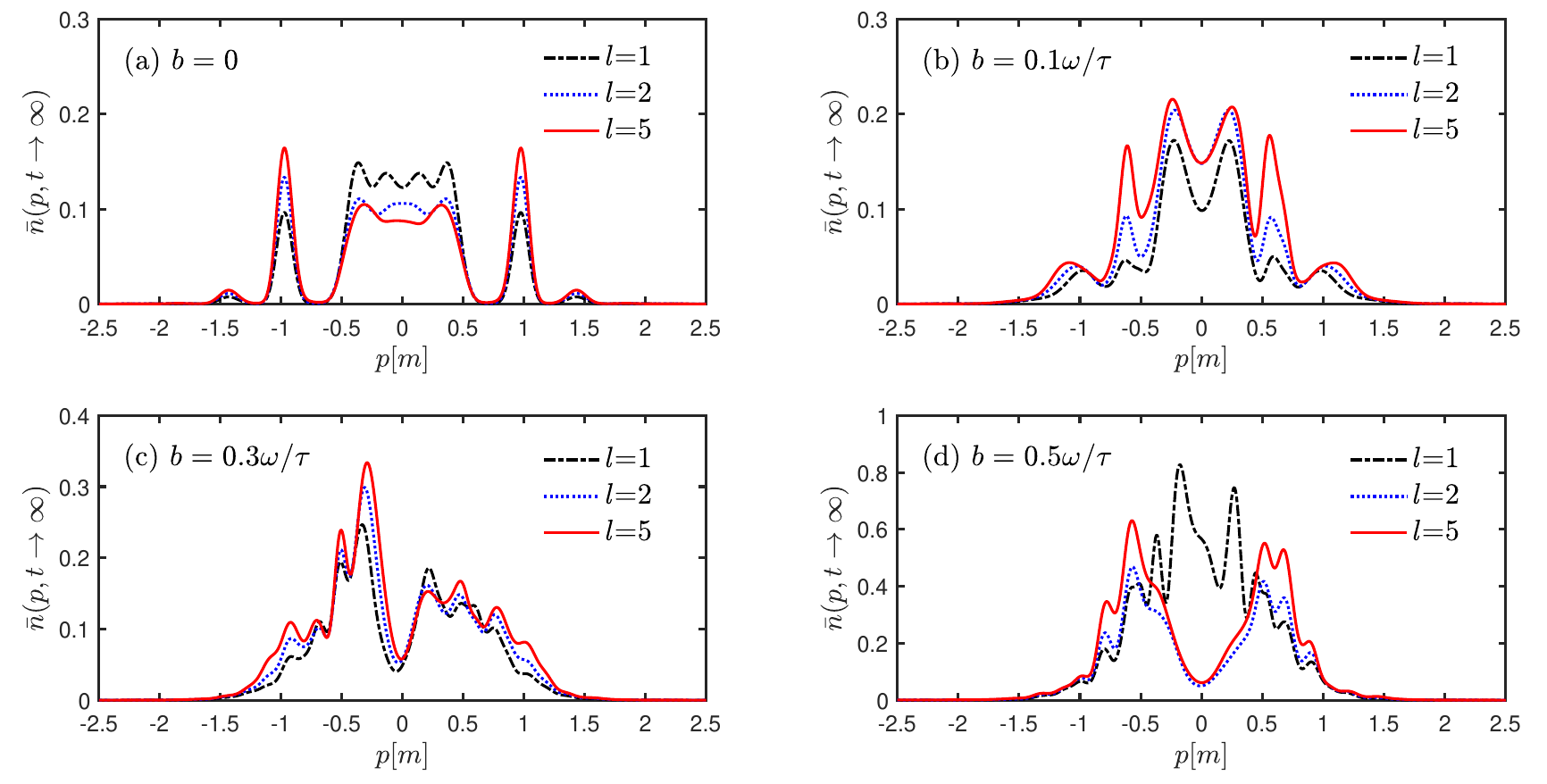}
\end{center}
\setlength{\abovecaptionskip}{-0.5cm}
\caption{Reduced momentum spectrum for different super-Gaussian envelope exponents in the high-frequency electric fields with $\lambda= 10~ \mathrm{m}^{-1}$. The other field parameters are the same as in Fig. \ref{fig:1}.}
\label{fig:2}
\end{figure}

Secondly, the momentum spectrum of the created particles for the finite spatial scale $\lambda =10 ~\mathrm{m}^{-1}$ is shown in Fig. \ref{fig:2}.
It is found that for vanishing chirp $b=0$, the dominant peak of the momentum spectrum becomes a broader compared with the peak in the quasi-homogeneous limit $\lambda =1000 ~\mathrm{m}^{-1}$, as shown in Fig. \ref{fig:2}(a), which may be associated with quantum interference. As known from atomic ionization, the multiphoton peak is interpreted as the result of particle trajectory superposition onto the observable interference pattern \cite{atomicionization:1993pb,atomicionization:2000gg,trajectories:2016BW}. However, the finite spatial scale of the laser field appears to weaken coherent superposition. In turn, the corresponding interference pattern is disrupted and the momentum distribution is broadened. Compared with the results of $l=1$, the width of the dominant peak for $l=2$ in the momentum spectrum narrows. This can be attributed to the shorter effective field interaction time at $l=2$, leading to a more concentrated momentum spatial distribution of pair production within the short time window. For $l=5$, the effective interaction time is even shorter, causing further narrowing of the main peak width.
With the introduction of the small chirp $b=0.1\omega/\tau$, the main peak splitting is observed in the momentum spectrum, see Fig. \ref{fig:2}(b). The peak splitting can be explained by ponderomotive effects in the high-frequency fields which have been reported in Ref. \cite{Kohlfurst:2017hbd}. Compared to the results of $l=1$, it is found that the oscillation amplitude for $l=2$ in the momentum spectrum is larger.
For $l=5$, the oscillation amplitude further increases due to changes in the time-domain characteristics of the electric fields.
However, for large chirp, the difference in the modulation of the momentum spectrum by the super-Gaussian envelope exponent $l$ also becomes indistinct, as shown in Figs. \ref{fig:2}(c) and \ref{fig:2}(d).
\begin{figure}[t]
\begin{center}
\includegraphics[width=\textwidth]{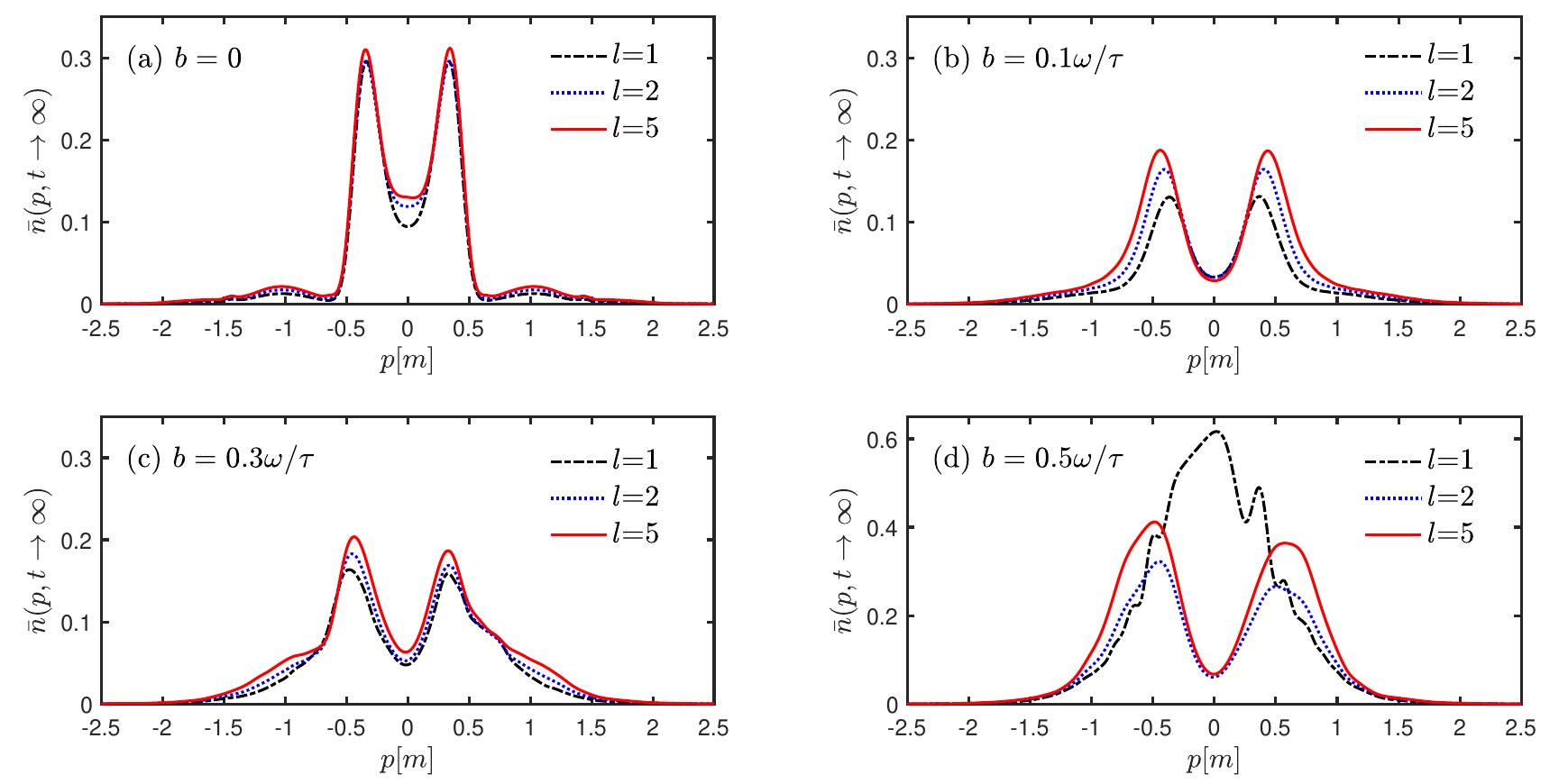}
\end{center}
\setlength{\abovecaptionskip}{-0.5cm}
\caption{Reduced momentum spectrum for different super-Gaussian envelope exponents in the high-frequency electric fields with $\lambda=2.5~ \mathrm{m}^{-1}$. The other field parameters are the same as in Fig. \ref{fig:1}.}
\label{fig:3}
\end{figure}

Finally, the momentum spectrum of created particles for the extremely small spatial scale $\lambda =2.5 ~\mathrm{m}^{-1}$ is shown in Fig. \ref{fig:3}. It is found that for vanishing chirp $b=0$, the middle peak splitting is observed on momentum spectrum, see Fig. \ref{fig:3}(a). The peak splitting can be also explained by ponderomotive force which pushes particles towards low field intensity regions in momentum space \cite{Kohlfurst:2017hbd}.
With the increasing super-Gaussian envelope exponent $l$ ($l=2,5$), the peak values in the momentum spectrum become higher and the valley values also lift.
The increase in the peak values can be explained as follows. When $l$ increases, the effective action time of the electric field is shortened. This enables particles to not have enough time to undergo a complete deceleration process after acquiring large momentum, which results in more particles "remaining" in the region of large momentum and thus increases the peak value in this region.
The lifting of the valley values can be understood as that the created particles with very small momentum are decelerated to zero by the external field, and because the super-Gaussian electric envelope field has a shorter action time than the usual Gaussian envelope field, which is insufficient to accelerate these particles, causing more particles to gather at zero momentum region.
As the increase of the chirp, momentum spectrum range widens, and the symmetry of the momentum spectrum is broken, resulting in the formation of two peaks with the left higher than the right in Figs. \ref{fig:3}(b) and \ref{fig:3}(c).
The reason for this result is that within the dominant time interval $-\tau \le t \le \tau$ of the pair production process, the external field oscillates slowly for $t \le 0$ and develops a broader temporal peak with increasing chirp. Therefore, the created particles do not have enough time to decelerate, leading to an increase in the particle number density at large momentum.
For large chirp $b=0.5\omega/\tau$, see Fig. \ref{fig:3}(d), it is found that peak splitting disappears at $l = 1$, which is associated with the highly nonuniform oscillations induced by the large chirp that suppress the formation of the ponderomotive force \cite{Ababekri:2019qiw}.
However, peak splitting persists for $l > 1$. It is assumed that this phenomenon may arise from the variation of the ponderomotive force caused by the combined effect between $l$ and $b$.
To make it easier to understand, a quantitative analysis will be conducted on the effects of the super-Gaussian envelope exponent $l$ on the ponderomotive force as follows.
It is well known that the ponderomotive force is governed by the effective mass gradient $\nabla_x m_*(x)$, i.e.,
\begin{equation}
F_p \propto -\nabla_x m_*(x),
\label{PondFf}
\end{equation}
with the effective mass defined as
 \begin{equation}\label{EffM}
\quad m_*(x) = m\sqrt{1 + \tilde{\xi}(x)^2}~~\mathrm{and}~\quad \tilde{\xi}(x) = \frac{e}{m} \sqrt{-\left\langle A_\mu(x,t)A^\mu(x,t)\right\rangle},
\end{equation}
where $\left\langle \cdot \right\rangle$ denotes the time average over the field oscillation period.
For simplicity, one may consider $\langle E^2(x,t) \rangle$ directly, because the ponderomotive force is also proportional to $\nabla \langle E^2(x,t) \rangle$. The time average \(\langle E^2(x,t) \rangle\) of Eq.(\ref{FieldMode}) can be separated into spatial and temporal parts as
\begin{equation}
\langle E^2(x,t) \rangle = E_0^2 \exp\left(-\frac{x^2}{\lambda^2}\right) \cdot S(l),
\label{PondFE}
\end{equation}
where $S(l)$ is the temporal average part as
\begin{equation}
S(l) = \frac{1}{\tau} \int_{-\tau}^{\tau} \exp\left(-\left(\frac{t}{\tau}\right)^{2l}\right) \cos^2(bt^2 + \omega t) dt.
\label{PondFEe}
\end{equation}
Thus, the ponderomotive force $F_p$ can be written as
\begin{equation}
F_p \propto -\nabla\left[ E_0^2 \exp\left(-\frac{x^2}{\lambda^2}\right) S(l) \right] = E_0^2 S(l) \cdot \frac{2x}{\lambda^2} \exp\left(-\frac{x^2}{\lambda^2}\right),
\label{PondFEeF}
\end{equation}
showing that the magnitude of $F_p$ is proportional to $S(l)$. Numerical calculation reveals that $S(l)$ increases with increasing $l$, which directly enhances $F_p$.
This is quantitatively manifested the increase in the ponderomotive force with the increasing $l$.
In contrast, previous work Ref. \cite{Ababekri:2019qiw} has shown that large chirp suppresses the formation of the ponderomotive force. The present analysis suggests that the super-Gaussian envelope exponent $l$ plays a dominant role when the chirp is sufficiently large. This explains the observed splitting of peaks in the momentum spectrum for $l > 1$, even in the presence of strong chirp.

\begin{figure}[t]
\begin{center}
\includegraphics[width=\textwidth]{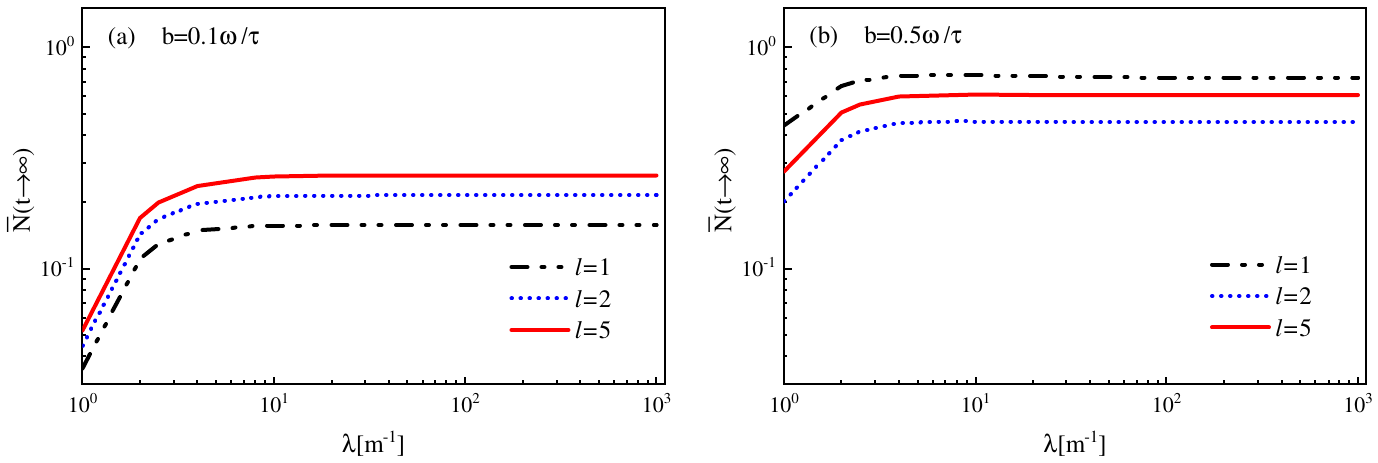}
\end{center}
\setlength{\abovecaptionskip}{-0.5cm}
\caption{Reduced total yield dependence on spatial scales for different super-Gaussian envelope exponents in the high-frequency electric fields. The chirp values are $b=0.1\omega/\tau$ for (a) and $b=0.5\omega/\tau$ for (b), respectively.}
\label{fig:4}
\end{figure}
In order to investigate the influence of the super-Gaussian envelope exponent more deeply, the reduced total yield of the created particles $\bar{N}(t \to \infty)$ with different super-Gaussian envelope exponents $l$ as a function of the spatial scales for different chirp is displayed in Fig. \ref{fig:4}.
From Fig. \ref{fig:4}(a), it can be seen that the total yield of the created particles increases with the spatial scale for any super-Gaussian envelope exponent $l$ ($l = 1, 2, 5$) when the small chirp $b=0.1\omega/\tau$.
Specifically, for small spatial scales, the total yield of the created particles rapidly decreases as the spatial scale is reduced. This is due to the fact that as the electric field energy decreases with a reduction in spatial scale, the total yield of particles created within the field region decreases correspondingly.
For large spatial scales, the total yield tends to a constant with the spatial scale increases.
Furthermore, for a fixed spatial scale, the total yield consistently increases as $l$ increases.
This phenomenon can be explained by the fact that an increase in the super-Gaussian envelope exponent $l$ narrows the temporal envelope, causing the electric field strength to reach its maximum rapidly, thereby enhancing the probability of pair production in the field.
From Fig. \ref{fig:4}(b),
it can be observed that the variation trend of the total yield of the created particles with the spatial scale is the same as that shown in Fig. \ref{fig:4}(a) for any super-Gaussian envelope exponent $l$ ($l = 1, 2, 5$) when the large chirp $b=0.5\omega/\tau$.
However, for a fixed spatial scale, the total yield reaches its maximum when $l=1$.
Of particular interest is that the total yield is relatively large for $l=5$, while it is smallest for $l=2$.

\subsection{Low Frequency Fields: $\omega=0.1~\mathrm{m}$}

In this subsection, it investigates the influence of the super-Gaussian envelope exponent $l$ on pair production in the low-frequency electric fields, and discusses the results for three spatial scales: $\lambda= 500~\mathrm{m}^{-1}$, $\lambda= 10~\mathrm{m}^{-1}$, and $\lambda= 2~\mathrm{m}^{-1}$.

\begin{figure}[t]
\begin{center}
\includegraphics[width=\textwidth]{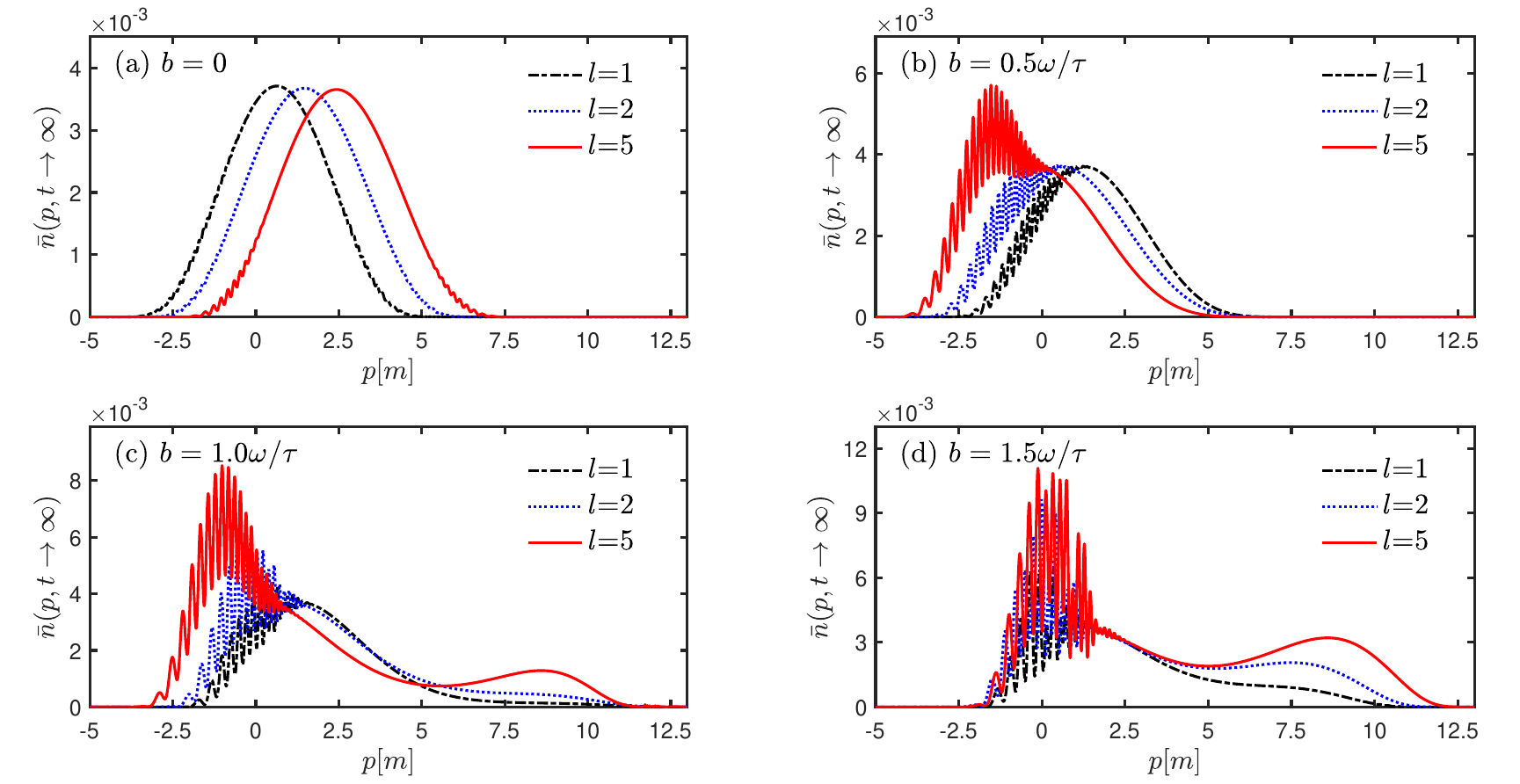}
\end{center}
\setlength{\abovecaptionskip}{-0.5cm}
\caption{Reduced momentum spectrum for different super-Gaussian envelope exponents in the low-frequency electric fields with $\lambda=500~\mathrm{m}^{-1}$. The other field parameters are $E_{0}=0.5E_\mathrm{cr}$, $\omega=0.1~\mathrm{m}$ and $\tau=25~\mathrm{m}^{-1}$.}
\label{fig:5}
\end{figure}

Firstly, the momentum spectrum for the quasi-homogeneous limit $\lambda=500~\mathrm{m}^{-1}$  is shown in Fig. \ref{fig:5}.
As shown in Fig. \ref{fig:5}(a), for vanishing chirp $b=0$, the center of the momentum spectrum does not corresponding to $p=0$, which is same to the homogeneous case of Fig. $2$ in the Ref. \cite{Hebenstreit:2009km}.
This is because the vector potential $A(t)\neq 0$ when the external field is switched off, so that $p_{x}=q-eA(t)$ takes a nonzero value $-A(t\rightarrow \infty)$, and this value corresponds to the momentum peak at the canonical momentum $q=0$.
To facilitate understanding, we have quantitatively computed the $p_{x}$-value for $l=1$.
Specifically, the spatial variation of the field can be neglected when $\lambda$ is sufficiently large, then
$A(t) = -\int_0^t E(t) \mathrm{d}t = -E_0 \int_0^t \exp\left[-\frac{1}{2}\left(\frac{t}{\tau}\right)^{2l}\right]\cos(\omega t) \mathrm{d}t$
can be obtained for $b=0$ as follows.
For $E_{0}=0.5 E_\mathrm{cr}$, $\omega=0.1\mathrm{m}$ and $\tau=25\mathrm{m}^{-1}$, we have $A(t\rightarrow \infty)\approx-0.693$ and the kinetic momentum $p_{x}\approx0.693$ at $l=1$.
This is approximately consistent with the numerical result $p_{x}\approx0.652$ at $l=1$ when $\lambda=500 \mathrm{m}^{-1}$, shown in Fig. \ref{fig:5}(a).
In other cases with different super-Gaussian envelop exponents $l$, the reason for the momentum peak shifting from zero is the same although the momentum values at the peak position are different because that $-A(t\rightarrow \infty)$ depends strongly on both $b$ and $l$.
As observed in Fig. \ref{fig:5}(a), it is also found that the momentum spectrum for $l=5$ exhibits weak oscillations. To understand this behavior,
we have depicted the structure of the corresponding turning points in Fig. \ref{fig:9}(a) and have provided a quantitative discussion in Sec. \ref{turningpoint}.
Furthermore, with the introduction of the small chirp $b = 0.5\omega/\tau$, stronger oscillations can be observed on the left side of the momentum spectrum for $l=1$, as shown in Fig. \ref{fig:5}(b). Compared with the results of $l=1$, more significant oscillations appear for $l=2$, and the momentum spectrum shifts towards the negative momentum direction. This trend is further enhanced when $l=5$. These oscillations can be understood as an interference effect of particles created by large peaks with opposite signs in the time field, and the increasing of $l$ enhances this effect.
Meanwhile, it is also observed that the momentum distribution range widens for large chirp in Figs. \ref{fig:5}(c) and \ref{fig:5}(d). At the same time, a secondary peak appears on the right-hand side, which is most pronounced for $l=5$. This is because the external field oscillation slows down significantly within the dominant time interval and causes particle acceleration.
\begin{figure}[t]
\begin{center}
\includegraphics[width=\textwidth]{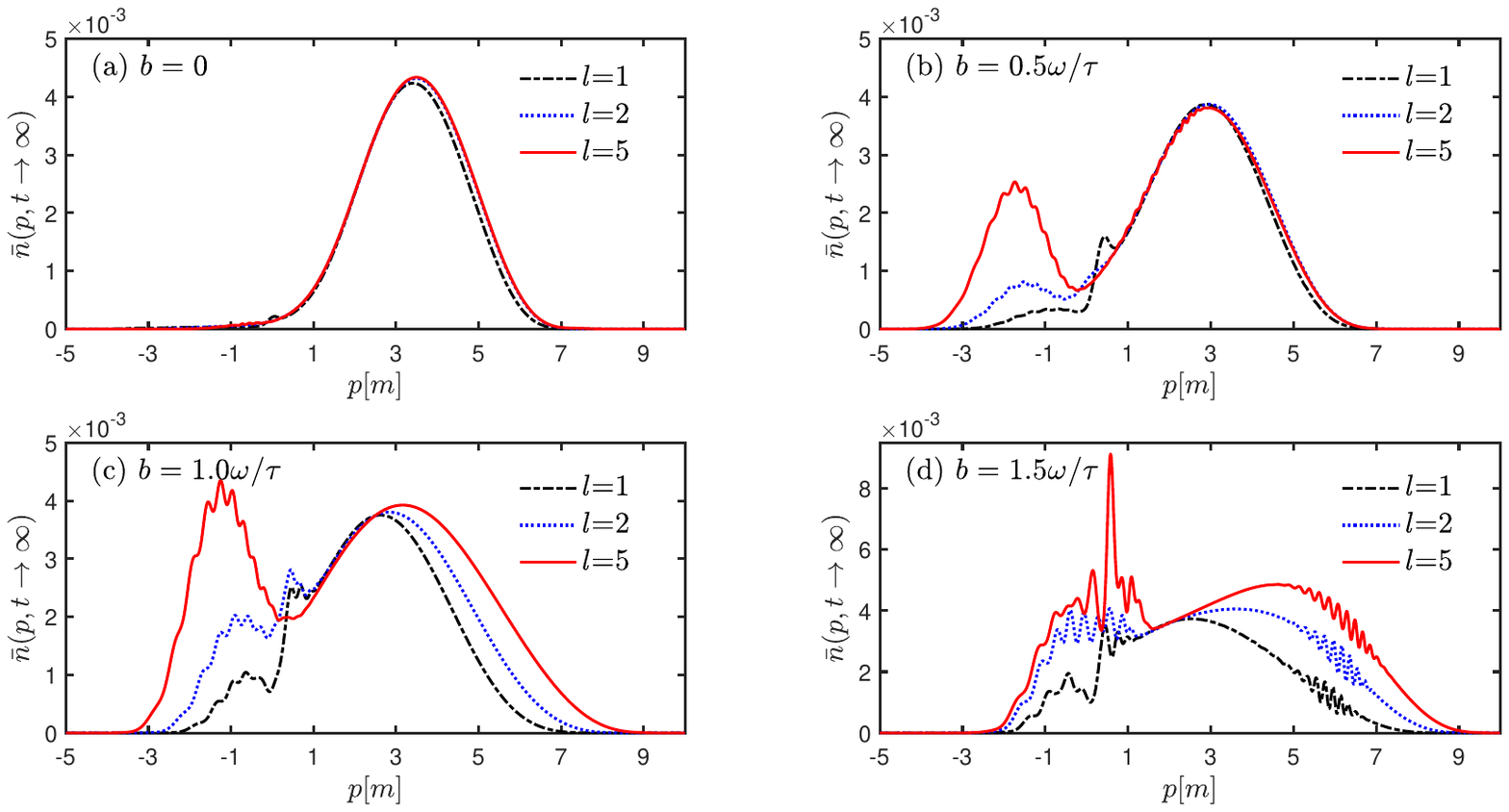}
\end{center}
\setlength{\abovecaptionskip}{-0.5cm}
\caption{Reduced momentum spectrum for different super-Gaussian envelope exponents in the low-frequency electric fields with $\lambda= 10~\mathrm{m}^{-1}$. Other field parameters are the same as in Fig. \ref{fig:5}.}
\label{fig:6}
\end{figure}

Secondly, the momentum spectrum for the finite spatial scale $\lambda =10 ~\mathrm{m}^{-1}$ is shown in Fig. \ref{fig:6}.
It can be seen that for vanishing chirp $b = 0$, the momentum spectrum is slightly wider for $l=2$ and $l=5$ compared with the result $l=1$, while the super-Gaussian envelope exponent $l$ hardly causes any significant change on the momentum spectrum peak, as shown in Fig. \ref{fig:6}(a).
For small chirp $b = 0.5\omega/\tau$, see Fig. \ref{fig:6}(b), weak oscillations can be observed on the left side at $l=1$ and the momentum spectrum broadens into negative momentum region. This broadening is because that in the present case of the decreasing spatial width, the particles created with a certain momentum leave the field region and miss the deceleration due to the negative field peak. For the large super-Gaussian envelope exponents ($l=2,5$), the peak in the negative momentum region increases further, and the reason of it is that the increase in $l$ shortens the effective interaction time of the external field, leading to a higher number of particles that miss deceleration.
For large chirp $b = 1.0\omega/\tau$ and $b = 1.5\omega/\tau$, as shown in Figs. \ref{fig:6}(c) and \ref{fig:6}(d), the momentum distribution range widens and shifts. This is because that the oscillation of the external field slows down greatly within the dominant time interval and causes particle acceleration.

\begin{figure}[t]
\begin{center}
\includegraphics[width=\textwidth]{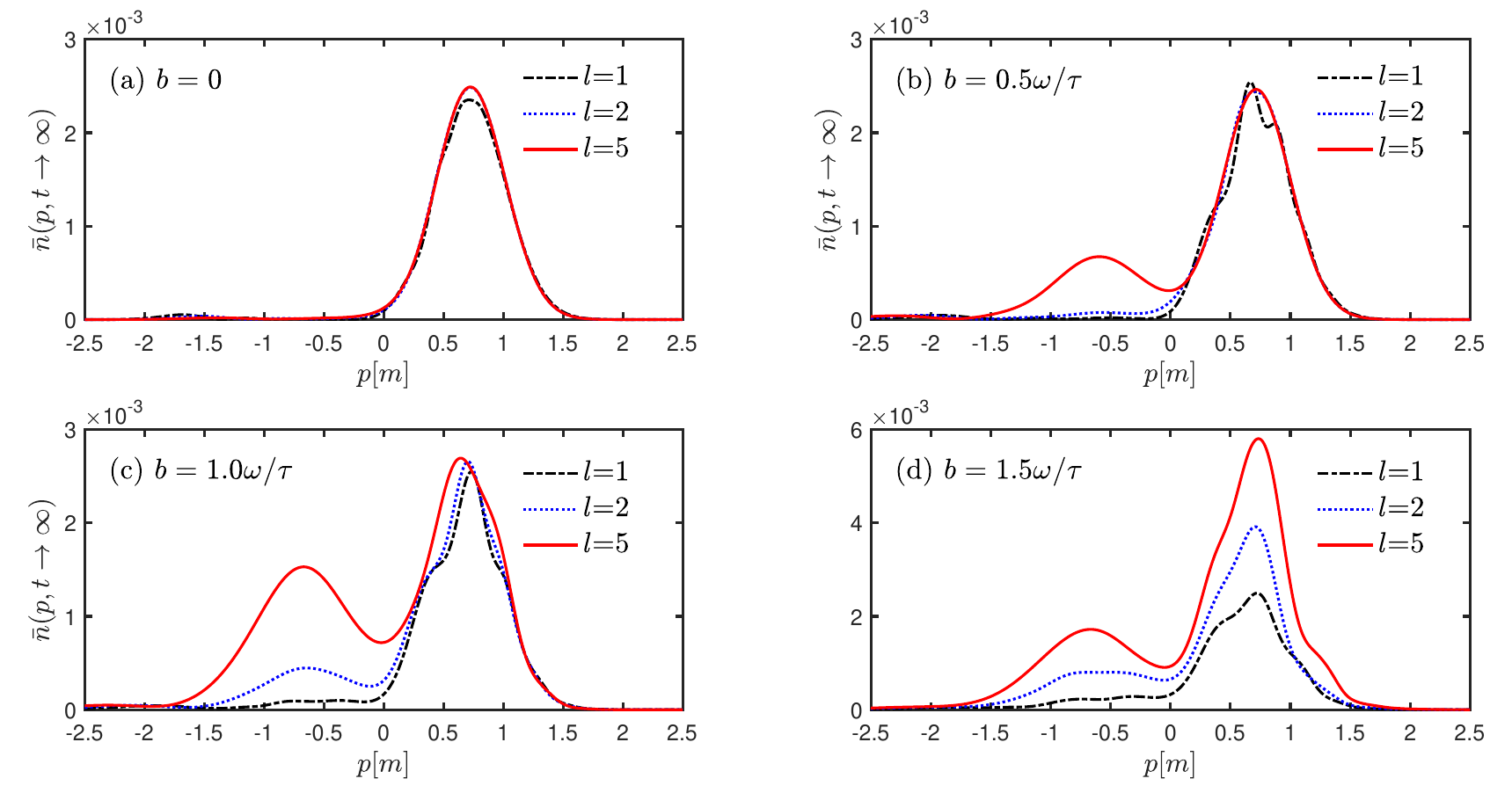}
\end{center}
\setlength{\abovecaptionskip}{-0.5cm}
\caption{Reduced momentum spectrum for different super-Gaussian envelope exponents in low-frequency electric fields with $\lambda= 2~\mathrm{m}^{-1}$. Other field parameters are the same as in Fig. \ref{fig:5}.}
\label{fig:7}
\end{figure}

Finally, the momentum spectrum for the extremely small spatial scale $\lambda =2 ~\mathrm{m}^{-1}$ is depicted in Fig. \ref{fig:7}. It is found that for vanishing chirp $b=0$, the variation of the super-Gaussian envelope exponent $l$ leads to a slightly increase in the peak value of the momentum spectrum, see Fig. \ref{fig:7}(a).
For small chirp $b = 0.5\omega/\tau$, it can be seen that no obvious oscillations are observed on the momentum spectrum for $l=1$ and the particle number density in the negative momentum region increases with the increase of $l$, as shown in Fig. \ref{fig:7}(b).
For large chirp $b = 1.0\omega/\tau$ and $b = 1.5\omega/\tau$, see Figs. \ref{fig:7}(c) and \ref{fig:7}(d), a weak oscillation appeared. Compared to $\lambda= 500 \mathrm{m}^{-1}$ and $\lambda= 10 \mathrm{m}^{-1}$, due to the extremely small spatial scale, the work done by the electric field is correspondingly diminished, resulting in the narrowest momentum spectrum width and the smallest peak value at $\lambda= 2 \mathrm{m}^{-1}$.

\begin{figure}[t]
\begin{center}
\includegraphics[width=\textwidth]{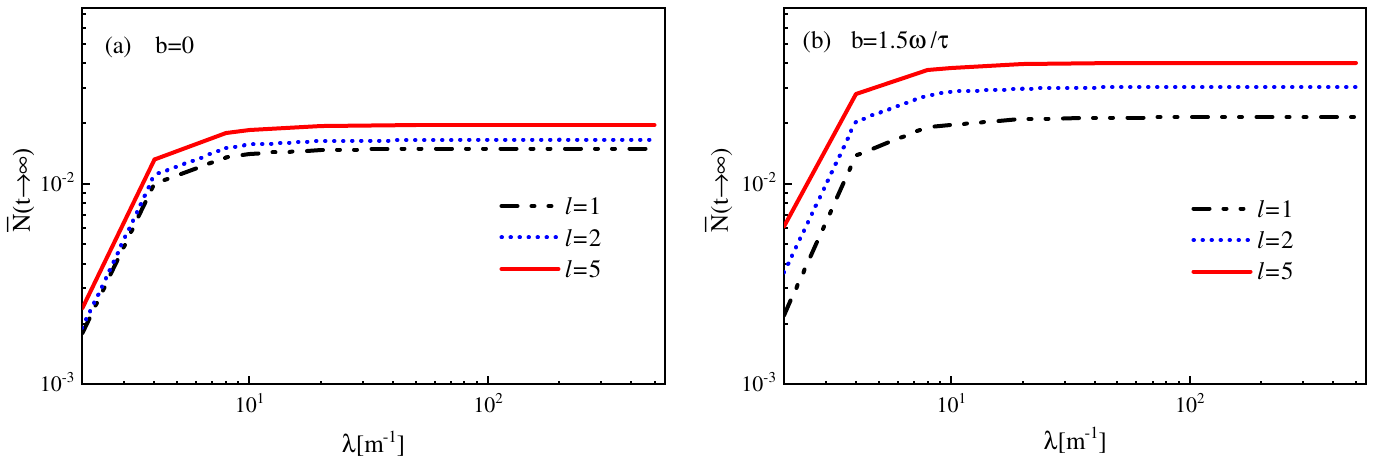}
\end{center}
\setlength{\abovecaptionskip}{-0.5cm}
\caption{The total yield dependence on spatial scales for the different super-Gaussian envelope exponents in the low-frequency electric fields. The chirp values are $b=0$ for (a) and $b=1.5\omega/\tau$ for (b), respectively.}
\label{fig:8}
\end{figure}

As shown in Fig. \ref{fig:8}, now let us see the reduced total yield of created particles with different super-Gaussian envelope exponents $l$ in a low-frequency field as a function of the spatial scale for different chirp.
It is found that the total yield of created particles varies differently with spatial scale for a fixed $l$.
At small spatial scales, the total yield of created particles decreases rapidly as the spatial scale shrinks.
However, at large spatial scales, the total yield of created particles tends to remain constant as the spatial scale increases. For low-frequency fields, when the chirp $b=0$, the total yield of created particles increases with $l$ at a fixed spatial scale, as seen in Fig. \ref{fig:8}(a).
More importantly, when the chirp reaches maximum value of $b=1.5\omega/\tau$, the increase in the total yield of the created particles becomes even more significant than results of $b=0$ and the total yield of created particles at $l=5$ is approximately two times that produced at $l=1$.
Overall, regardless of the chirp variations, the total yield eventually reaches its maximum as super-Gaussian envelope exponents $l$ increases.

\section{Semiclassical analysis and discussion}\label{turningpoint}

To gain deeper insight into some of the new features appearing in the momentum spectrum of the created particles, the WKB scattering approach can be used as an analytic way to qualitatively discuss the corresponding results \cite{Dumlu:2010vv,Dumlu:2011rr,NonrelativisticTheory:2003,Phase-Integral:1962}.
The turning points are obtained by solving the equation $\Omega_{\bm p}(t)= \sqrt{m^{2}+p_{\perp}^{2}+[p_{x}-eA(t)]^{2}}=0$, and these points located in the complex $t$-plane as complex-conjugate pairs \cite{Dumlu:2010vv}.

The approximate expression of multiple turning points for the pair production rate can be given by \cite{Dumlu:2011rr}
\begin{equation}\label{AE}
\begin{aligned}
N\approx \sum\limits_{i=1}^{n}e^{-2K_{i}} - \sum\limits_{{i}\neq{j}} 2\cos(2\theta_{(i,j)})
e^{-K_{i}-K_{j}},
\end{aligned}
\end{equation}
with
$$ K_{i}=\left\vert \int_{t_{i}^{*}}^{t_{i}}{\Omega_{\bm p}(t)dt} \right\vert $$
and
$$\theta_{(i,j)}=\left\vert\int_{Re (t_{i})}^{Re(t_{j})}{\Omega_{\bm p}(t)dt} \right\vert,$$
where $t_{i}$ and $t_{j}$ denote different turning points, and Re($t_{i}$) and Re($t_{j}$) denote the real part of the turning point.
It is well known that in the turning point structure, the closest turning points to the real $t$-axis have the dominant contribution to pair production rate due to the first term in Eq.(\ref{AE}).
The interference term $\theta_{(i,j)}$ present in the second term of Eq.(\ref{AE}) represents an integral between different turning points and accounts for the oscillatory behavior observed in the momentum spectrum.
Due to the similar physical reasons, this section only focuses on discussing two cases of a high-frequency field with large chirp $b=0.5\omega/\tau$ and a low-frequency field with vanishing chirp $b=0$.
\begin{figure}[t]
\begin{center}
\includegraphics[width=\textwidth]{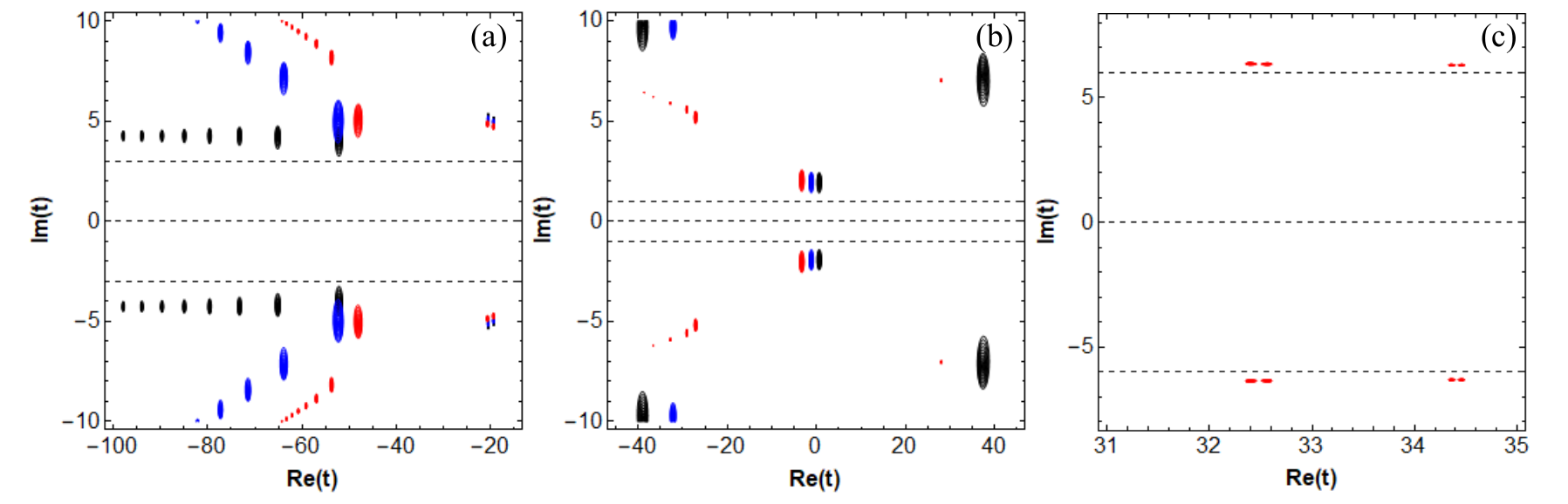}
\end{center}
\setlength{\abovecaptionskip}{-0.5cm}
\caption{Contour plots of $|\Omega_{\bm p}(t)|^{2}$ in the complex$t$ plane, showing the location of turning points distribution where $\Omega_{\bm p}(t)=0$. The black, blue, and red dots represent the results for $l=1,2$ and $5$, respectively. Panel (a) is the case for high-frequency field with chirp $b=0.5\omega/\tau$. Panel (b) is the case of the low-frequency field with vanishing chirp $b=0$. Panel (c) is the amplified plot of turning points at around Re(t) $\sim 33$ in panel (b). The three dashed lines are for eyes guidance.}
\label{fig:9}
\end{figure}

The structure of the turning points for a high-frequency field with chirp $b=0.5\omega/\tau$ is shown in Fig. \ref{fig:9}(a). Clearly, the turning points for $l=1$ (black dots) are located closest to the real $t$ axis compared to the results for $l=2$ (blue dots) and $l=5$ (red dots).
It is well-known that the closer the dominant turning point lies in the real $t$-axis, the greater the particle number density.
Accordingly, the particle number density $\bar n \left(\left(p_\perp = 0, p_{x}=0\right) , t\rightarrow \infty \right)\approx 0.53$ is the largest when $l=1$ in Fig. \ref{fig:1}(d).
Meanwhile, for $l=2$ and $l=5$, the main turning point pairs are located at nearly the same distance from the real $t$-axis. Correspondingly, the particle number densities are approximately equal, $\textit{i.e.}$
$\bar n \left(\left(p_\perp = 0, p_{x}=0\right) , t\rightarrow \infty \right)\approx 0.045$ for $l=2$, and $\bar n \left(\left(p_\perp = 0, p_{x}=0\right) , t\rightarrow \infty \right)\approx 0.054$ for $l=5$, respectively, as shown in Fig. \ref{fig:1}(d). Thus, the structure of the turning points in Fig. \ref{fig:9}(a) can be used to qualitatively explain the feature observed at $p_{x}=0$ in the momentum spectrum shown in Fig. \ref{fig:1}(d).

As shown in Fig. \ref{fig:9}(b), it can be observed that in the case of a low-frequency field with vanishing chirp $b=0$, the black points for $l=1$, blue points for $l=2$, and red points for $l=5$ at the center of the figure are progressively farther from the real $t$-axis as $l$ increases. The corresponding particle number densities are $\bar n \left(\left(p_\perp = 0, p_{x}=-1\right) , t\rightarrow \infty \right)\approx 2.25 \times 10^{-3}$ ($l=1$), $\bar n \left(\left(p_\perp = 0, p_{x}=-1\right) , t\rightarrow \infty \right)\approx 1.12 \times 10^{-3}$ ($l=2$), and $\bar n \left(\left(p_\perp = 0, p_{x}=-1\right) , t\rightarrow \infty \right)\approx 2.68 \times 10^{-4}$ ($l=5$) in Fig. \ref{fig:5}(a), exhibiting a decreasing trend.
This result can be also used to explained the particle number density reaches maximum at $p_{x}=-1$ for $l=1$ compared with the other cases ($l=2$ and $l=5$) in the momentum spectrum shown in Fig. \ref{fig:5}(a).
Meanwhile, for $l=5$ (red dots), it is evident that some complex conjugate pairs of turning points with nearly equal distances to the real axis appear in the right half of the real $t$-region, which is clearly visible in Fig. \ref{fig:9}(c). It is well known that interference effects are determined by the different turning point pairs that are located at nearly equal distances from the real axis. Thus, a slight oscillatory feature appears in the momentum spectrum for $l=5$, as shown in Fig. \ref{fig:5}(a).

\section{Conclusion and outlook}\label{conclusion}

In this work, the effects of super-Gaussian pulse shapes on pair production in the spatially inhomogeneous chirped electric fields is investigated using the DHW formalism, and the reduced momentum spectrum and the reduced total yield of created particles are mainly studied in the high- and low-frequency fields within our studied range of the super-Gaussian envelope exponent $l$ ($l=1,2,5$). Moreover, the semiclassical WKB approach and the corresponding turning points structure are employed to make some qualitative discussions. Our results are summarized briefly as follows.

For the high-frequency fields, it is found that the momentum spectrum exhibits pronounced oscillations with the variation of the super-Gaussian envelope exponent, and the total yield of created particles increases monotonically with increasing the super-Gaussian envelope exponent for small chirp and reaches a maximum at $l=5$ but varies non-monotonically for large chirp and reaches a maximum at $l=1$.
For the low-frequency fields, the momentum spectrum shifts and broadens as the super-Gaussian envelope exponent increases, and the total yield of created particles also increases monotonically with increasing the super-Gaussian envelope exponent for any chirp, meanwhile, the total yield under the super-Gaussian pulse electric field for large chirp is approximately twice that produced with the usual Gaussian pulse envelope.

These results demonstrate that the super-Gaussian pulse shape has a significant modulation effect on pair production, and the combined effect of the super-Gaussian pulse shape, frequency chirp, and spatial inhomogeneity not only gives rise to rich features in the momentum spectra but also leads to an enhancement of the total yield, which provides theoretical guidance for optimizing the form of external field to enhance the vacuum pair production and gives more reliable predictions for future experiments.

\section{Acknowledgments}

This work is supported by the National Natural Science Foundation of China (NSFC) under Grants No. 12565023, No. 12535015, No. 12375240 and No. 12165018, the Youth Science and Technology Foundation of Gansu Province under Grant No. 24JRRA276, and the Gansu Science and Technology Program under Grant No. 23JRRA681.


\end{document}